\documentstyle[11pt,newpasp,twoside,psfig,epsf]{book}

\newcommand{\eg}{{\em e.g.},\ }

\newcommand{\etc}{{\em etc.}\ }

\begin{document}

\pagenumbering{arabic}
\setcounter{page}{353}

%\documentstyle[12pt]{article}
%\oddsidemargin 0in
%\evensidemargin 0in
%\topmargin -0.5in  % might need to modify this for local versions of LaTeX
%\headheight .125in
%\footheight .125in
%\textheight 9.0in
%\textwidth 6.5in
%\parindent 0in  % Paragraph indentation amount
%\parskip 1ex    % White space between paragraphs amount
%\clubpenalty 100000  \widowpenalty 100000
%\nofiles

\setcounter{section}{0}
\setcounter{figure}{0}
\setcounter{table}{0}
\setcounter{footnote}{0}

%\begin{document}
%\pagestyle{plain}
%\thispagestyle{empty}

\markboth{NVO White Paper}{NVO White Paper}

\title{Toward a National Virtual Observatory: Science Goals, 
Technical Challenges, and Implementation Plan\footnotemark}
\footnotetext{This draft white paper was prepared by the informal 
NVO interim steering committee, whose members include T. Boroson
(NOAO), R. Brissenden (CXC), R.J. Brunner (CIT), T. Cornwell (NRAO),
D. De Young (NOAO), S.G. Djorgovski (CIT), R. Hanisch (STScI),
G. Helou (IPAC), C. Lonsdale (IPAC), T. Prince (CIT), E. Schreier
(STScI), S. Strom (NOAO), A. Szalay (JHU), D. Tody (NOAO), and N.White
(HEASARC). This version of the white paper has been modified by the
editors to conform to the style guidelines for these proceedings, and
is included here by agreement of the NVO interim steering
committee. The original version, dated June 8, 2000, is available from
the conference web site: http://astro.caltech.edu/nvoconf/.}

\begin{abstract}
The National Academy of Science Astronomy and Astrophysics Survey
Committee, in its new Decadal survey entitled {\em Astronomy and
Astrophysics in the New Millennium}, recommends, as a first priority,
the establishment of a National Virtual Observatory.  The NVO would
link the archival data sets of space- and ground-based observatories,
the catalogs of multi-wavelength surveys, and the computational
resources necessary to support comparison and cross-correlation among
these resources.  This White Paper describes the scientific
opportunities and technical challenges of an NVO, and lays out an
implementation strategy aimed at realizing the goals of the NVO in
cost-effective manner.  The NVO will depend on inter-agency
cooperation, distributed development, and distributed operations.  It
will challenge the astronomical community, yet provide new
opportunities for scientific discovery that were unimaginable just a
few years ago.
\end{abstract}

\section{Executive Summary}

Technological advances in telescope and instrument design during
the last ten years, coupled with the exponential increase in 
computer and communications capability, have caused a {\em dramatic
and irreversible change in the character of astronomical research.}
Large scale surveys of the sky from space and ground are being 
initiated at wavelengths from radio to X-ray, thereby generating
vast amounts of high quality irreplaceable data.

{\em The potential for scientific discovery afforded by these new surveys
is enormous.}  Entirely new and unexpected scientific results of major
significance will emerge from the combined use of the resulting datasets,
science that would not be possible from such sets used singly.
However, their large size and complexity require tools and structures
to discover the complex phenomena encoded within them.  We propose
establishing a {\bf National Virtual Observatory} that can meet these needs
through the coordination of diverse efforts already in existence as
well as providing focus for the development of capabilities that do
not yet exist.  The NVO will act as an enabling and coordinating entity
to foster the development of tools, protocols, and collaborations necessary
to realize the full scientific potential of astronomical databases in
the coming decade.  When fully implemented, the NVO will serve as 
{\em an engine of discovery for astronomy.}

The new scientific capabilities that will be enabled by the NVO are
essential to realize the full value of the terabyte and petabyte
datasets that are in hand or soon to be created.
Rapid querying of large scale catalogs,
establishment of statistical correlations, discovery of new data
patterns and temporal variations, and confrontation with sophisticated
numerical simulations are all avenues for new science that will be
made possible through the NVO.  In addition, the
NVO and its data archives will require
{\em collaborations with the computer science community,}
will provide opportunity for collaboration
with other disciplines facing similar challenges,
and will be 
{\em a venue for educational outreach.}
Three examples of scientific
programs involving Active Galactic Nuclei, the Large Scale Structure of
the Universe, and the structure of our Galaxy illustrate the scientific
promise of the NVO.
{\em The NVO will be technology-enabled, but science-driven.}

Implementation of the NVO involves significant technical challenges.
These include both the incorporation of existing data archiving efforts
in astronomy as well as the development of new capabilities and
structures.  Major technical components to the NVO include archives,
metadata standards, a data access layer, query and computing services,
and data mining applications.  Development of these capabilities will
require close interaction and collaboration with the
information technology community.

The implementation plan for the NVO is defined by four phases, beginning
with activities initiated prior to the establishment of the NVO and
leading to the fully operational phase of the NVO four--five years after its
inception.  This implementation plan is designed to begin placement of
deliverables and capability to the community at the earliest possible
time. This early functionality is essential to the success of the NVO.

\section{Introduction: Winds of Change}

For over two hundred years, the usual mode of carrying out
astronomical research has been that of a single astronomer or small group
of astronomers performing observations of a small number of objects. 
In the past, entire careers have been spent in the acquisition of enough data
to barely enable statistically significant conclusions to be
drawn.  Moreover, because observing time with the most
powerful facilities is very limited, 
many astrophysical questions that require a large amount of
data for their resolution simply could not be addressed. 

This approach is now undergoing a dramatic and very
rapid change.  The transformation is being driven by technological 
developments over the last decade that are without precedent.  The 
major areas of change upon which this revolution in astronomy
rests are advances in telescope design and fabrication, the development of
large scale detector arrays, the exponential growth of computing capability,
and the ever expanding
coverage and capacity of communications networks.  

The advances in telescope
design and fabrication have made possible the great space based observatories, 
opening new vistas in
gamma ray, X-ray, optical and infrared astronomy.  Advanced technology has
also made possible the establishment of a new generation of large aperture
ground based optical and IR telescopes as well as the design and construction
of single dish and multi element arrays operating at millimeter and centimeter
wavelengths.  At optical and infrared wavelengths these advances have been
coupled with the development of extremely sensitive, high resolution detector
arrays of ever increasing size, and the ability to mosaic these arrays has
resulted in instruments with fields of view of order 30 arcminutes 
and with $\sim 10^{8}$ 
pixels per image.  These technical developments continue to mature, with
more sophisticated and larger aperture telescopes being planned in space
and on the ground, using ever larger and more capable arrays of detectors
incorporated into advanced instrumentation.  Just as Moore's Law reflects
the exponential increase in computing capability with time, the technological
developments in observational astronomy over the last decade have in effect
introduced {\em a Moore's Law for astronomy as well.}

The emergence of more astronomical facilities, on the ground and in
space, with larger apertures and more sophisticated instrumentation,
will have a critical and inevitable consequence: an enormously increased
flow of data.  For example, the current data production rate of HST is
about 5 Gigabytes per day; but a facility recently recommended for
construction by the AASC Decadal survey---the Large-Aperture Synoptic
Survey Telescope---could produce up to 10 terabytes per day!

In addition to this increased data rate, 
{\em the manner in which observations are being made is also changing.}
Although the new observatories in
space and on the ground still devote a significant fraction of their
time to research in the ``single observer/single program'' mode where
small blocks of time are allocated to many 
specifically targeted research programs, 
more time is now being devoted to large scale surveys
of the sky, often at multiple wavelengths, that  
involve large numbers of collaborators.

These large survey programs will produce coherent blocks of data
obtained with uniform standards, and with the
amount of data often measured in terabytes.   This {\em paradigm shift}
has been made possible not only by the increased capabilities
of the new facilities that permit much faster acquisition of data,
but also by the availability of computational hardware and software that 
make it possible to acquire, reduce, and archive this data. 

A major technological development that will change the character of
astronomical research is the advent of high speed information transfer
networks with broad coverage.  Although the rapid transfer of large
amounts of data over common networks is currently unacceptably 
slow (over 20 {\it days} to transfer a 1 terabyte data set), 
future networks will be much faster.
The availability of these data rates, together with the high efficiency
of data acquisition at both ground and space based facilities,
will make possible the efficient transmission of large amounts of data
to many different sites.  This technology will also enable access to
specific subsets of data by an extensive user community that prior to
this had no readily available access to these data; the potential
scientific yield resulting from this accessibility will be enormous.   

It is clear that all of these technological drivers will result in
{\em an unprecedented flow of astronomical data} in the coming years.
Moreover, {\em these data sets will be very different,} 
in that most of them
will be in the form of coherent surveys, often at multiple wavelengths,
over significant portions of the sky.  Hence they will have 
{\em a richness and depth that is unprecedented,} 
and they will present unique opportunities 
for application to a variety of scientific programs by a wide range of
users.  This aspect alone makes the systematic archiving of these data
essential.  In addition, the data will be obtained through use of costly
state of the art facilities that will be highly oversubscribed, and this
will essentially preclude repetitions of observations previously made;
this also argues for a general archiving of these data.  

The existence
of such archives, containing multiwavelength data on hundreds of millions
of objects, will clearly create a demand within the astronomical community
for access to the archives and for the tools necessary to analyze the data
they contain.  Opportunities for data mining, for sophisticated pattern
recognition, for large scale statistical cross correlations, and for the
discovery of rare objects and temporal variations all become apparent.

In addition, for the first time in the history of astronomy, such data
sets will allow meaningful comparisons to be made between sophisticated
numerical simulations and statistically complete multivariate bodies of 
data. The rapid growth of high speed and widely distributed networks means
that all of these scientific endeavors will be made available to the
community of astronomers throughout the US and in other countries.

These technological developments have converged in the last few years, 
and they will 
{\em completely alter the manner in which most observational astronomy
is carried out.}  These changes are inevitable and irreversible, and they
will have dramatic effects on the sociology of astronomy itself. 
Moreover, there is a growing awareness, both in this country and
abroad, that the acquisition, organization, analysis and dissemination
of scientific data are essential elements to {\em a continuing robust
growth of science and technology.}  These factors make it imperative to
provide a structure that will enable the most efficient and
effective synthesis of these technological capabilities.  Hence there is a 
need now for an entity such as a National Virtual Observatory to
oversee the rational disposition of the growing body of astronomical
data. 

\section{The Vision of a National Virtual Observatory} 

\subsection{Structure and Function: Enabling New Science} 

The NVO is an entity that will enable advances in astronomy and
astrophysics previously unattainable.
It will be a key ingredient in establishing
a new Age of Discovery in astronomy.
With its conjugation of terabyte data archives, image libraries of
millions of objects at wavelengths from gamma rays to radio frequencies,
sophisticated data mining and analysis tools, access to terascale
supercomputing facilities with petabyte storage capacities, and very
high speed connectivity among major astronomical centers, the NVO will
be unique.  It will 
make possible rapid querying of individual terabyte archives by 
thousands of researchers, enable visualization 
of multivariate patterns embedded in large catalog and image databases, 
enhance discovery of complex patterns or rare phenomena, encourage
real time collaborations among multiple research groups, and allow
large statistical studies that will for the first time permit
confrontation between databases and sophisticated numerical simulations. 
It will facilitate our understanding of many of the astrophysical 
processes that determine the evolution of the Universe.  It will enable
new science, better science, and more cost effective science. 
The NVO will act as a coordinating and enabling
entity to foster the development of tools, protocols, and collaborations
necessary to realize the full scientific potential of astronomical
databases in the coming decade.  {\em It will be an engine of discovery for
astronomy}.  

To accomplish this, the NVO will first and foremost be built as a science
driven, community effort with a major fraction of the funding disbursed via
a peer review process.  This would be accomplished through regular
announcements of opportunity for both software projects that develop NVO
infrastructure and for science activities that utilize the NVO.  More
specifically, NVO activities to fulfill its role would include:
\begin{itemize} 
\item
Establishment of a common systems approach to data
pipelining, archiving and retrieval that will ensure easy access by a
large and diverse community of users and that will minimize costs and times
to completion; 
\item
Enabling the distributed development of a suite of
commonly usable new software tools to make possible the querying,
correlation, visualization and statistical comparisons described above;
\item
Coordinating the establishment of high speed data transfer networks
that are essential to providing the connectivity among archives, terascale
computing facilities, and the widespread community of users; 
\item
Facilitating productive collaborations among astronomy centers and 
major academic institutions, both national and international, in order
to maximize productivity and minimize infrastructure costs; 
\item
Ensuring communication and possible collaborations
with scientists in other disciplines facing similar problems, and 
with the private sector;
\item
Maintaining a continuing program of public and educational outreach
that capitalizes upon the unique resources, in both data and software,
of the NVO to provide a unique window into astronomy and scientific
methodology. 
\end{itemize}

\subsection{Design Philosophies}

The NVO will be a unique entity, primarily because its operation will be
distributed and will be based upon rapidly developing technologies in 
communication and computer science. In order to ensure its continuing
vitality, the NVO must embrace several major themes.  
\begin{itemize}
\item
The NVO must be \textbf{evolutionary}.  From its inception, this 
evolutionary nature and will enable the NVO to  
respond quickly to changing technical and scientific opportunities 
and community requirements.  Because of the continuing evolution
in technical capabilities, this evolutionary nature will be an 
integral part of the NVO throughout its existence.
An immediate consequence of this agility is the need for a
management structure that both manages the distributed development
efforts of the NVO and that moves quickly to exploit new possibilities.
Management and
oversight must be effective, efficient, visionary, and accountable to the
community, yet minimize overhead and inertia.
\item
The NVO must be \textbf{distributed} in nature.  Significant amounts
of expertise are already in place at existing centers, and full advantage
will be taken of this from the outset.  In addition, the most economical
and effective progress toward the goals of the NVO may well be realized
in its operational phase through a distributed approach.  This would
entail location at existing centers and at future data centers those key
areas of NVO functions that are most effectively carried out at those
centers.
\item
The NVO must be \textbf{integrated}. 
Complementary to its distributed structure will
be an enduring theme of integration as the NVO evolves.  In order for 
the NVO to be most effective in facilitating scientific advances, the 
information technology functions must be integrated over all wavelengths 
and over space based and ground based facilities.  In addition, integration
with developments in computer science and information technology will be
an essential element of the NVO. 
\item
The NVO must provide \textbf{outreach}.
The vast datasets and the accompanying analysis tools that will be
available through the NVO provide an opportunity for educational and
public outreach on a level that has not been possible in the past.
An active outreach program that takes full advantage of the educational
potential of the NVO resources must be implemented at all stages of the
NVO development.
\item 
The NVO must be \textbf{globally oriented}. 
A continuing aspect of the NVO 
will be its international links to similar efforts in other countries.
Though it will not initially be an international collaboration itself, it
is clear that the NVO must maintain communication, and collaborations 
when appropriate, at all levels with these other activities.  
It seems inevitable that NVO engendered activities will become a worldwide
phenomenon.  
\item 
The NVO must \textbf{provide a path to the future}.
The vision of NVO described here is primarily that of
a catalytic and enabling entity, with minimal structure and enormous
connectivity.  A direct product of this will be the
enhancement of scientific productivity in astronomy to new levels.  However,
a larger and perhaps more enduring legacy of the NVO will be its role in the
establishment of an \textbf{astronomy information infrastructure} 
within the US and
throughout the world.  The growth of this infrastructure, expedited by the 
NVO, will provide unprecedented new vistas and opportunities for 
astronomical research in the future.
\end{itemize} 

\section{The Scientific Case for the NVO}  

As we look ahead, the astronomical community stands poised to take advantage 
of the breathtaking advances in 
computational speed, storage media and detector technology in two ways: (1) 
by carrying out new generation surveys spanning a wide range of 
wavelengths and optimized to exploit these advances fully; and (2) by 
developing the software tools to enable discovery of new patterns in the 
multi-terabyte (and later petabyte) databases that represent the legacies 
of these surveys. In combination, new generation surveys and software 
tools can provide the basis for enabling science of a qualitatively 
different nature. 

Moreover, {\em the inherent richness of these databases promises scientific 
returns reaching far beyond the primary objectives of the survey:} 
for example, repeated imaging surveys aimed at developing a census of Kuiper 
Belt objects can provide the basis for discovering supernovae at $z > 1$. 
Indeed, the multiplier effects of survey databases are enormous as 
exemplified by the world-wide explosion of research ignited by the Hubble 
Deep Field. 

We now have the tools to carry out surveys over nearly the entire 
electromagnetic spectrum on a variety of spatial scales and over multiple 
epochs, all with well-defined selection criteria and well-understood limits.  
The ability to create panchromatic images, and in some cases digital movies 
of the universe, provide 
{\em unprecedented opportunities for discovering new 
phenomena and patterns that can fundamentally alter our understanding.} 
In the 
past, a panchromatic view of the same region of sky at optical and radio 
wavelengths led to the discovery of quasars. The availability of infrared data 
led to the discovery of obscured active galactic nuclei and star-forming 
regions unsuspected from visible images. Repeated images of the sky have led 
to the discovery of transient phenomena---supernovae, and more recently, 
micro-lensing events---as well as deeper understanding of synoptic phenomena.
The joining together of various large scale digital surveys will make  
possible new explorations of parameter space, such as the low surface 
brightness universe at all wavelengths.

{\em The challenges of discovering new patterns and phenomena in huge 
astronomical databases find parallels in the medical, biological and earth 
sciences}.  For example, the size of the human genome is roughly 3 GB, 
while a digital all sky survey will be about 10 TB in size.
The development of tools and techniques to handle astronomical datasets of
this size will clearly have to call upon new developments in computer
science and will, when developed, have applications to fields outside
astronomy.  In all cases, the full power of these
databases cannot be tapped without the development of new tools and new
institutional structures that can consolidate disparate databases and
catalogs, enable access to them, and place analysis tools in the hands of
a broad community of imaginative scientists. It is this vision that
motivates the creation of the NVO.

The major capabilities of the NVO that need to be established
in order to enable its scientific goals include the ability to:

\begin{itemize}
\item
Federate existing large databases at multiple wavelengths and create tools to 
query them in both the catalog and the pixel domain;
\item
Develop universal standards for archiving future large databases;
\item
Provide a framework for incorporating new databases, thus minimizing the 
cost of new surveys and experiments and maximizing their scientific return; 
\item 
Develop analysis tools for discovery in catalog datasets and for statistical 
analysis of resulting joint datasets;
\item
Develop tools for object classification in the image datasets;
\item
Develop tools for visualization in both catalog and image databases;
\item
Develop new approaches to querying image databases and for image analysis 
and pattern recognition;
\item
Incorporate the results of sophisticated numerical simulations and develop a 
statistical ``toolbox'' for confronting these simulations with data; and
\item 
Link with existing and future digital libraries and journals.
\end{itemize} 

All of the above are possible, and all are qualitatively different from what we 
now do because of size, dimensionality, and complexity. Over time, the NVO can 
evolve to carry out all these functions. 
However, while enabled by technology, the NVO is not driven by 
technology. Instead, its structure and evolution are fundamentally driven by 
science and the needs of the scientific community. 
A major tool for guiding decisions about 
developing an NVO capability and the pace of 
implementing these capabilities will be that of a community-developed 
``Science Reference Mission'' (SRM) for NVO.
We envision a structured process to develop the SRM for NVO comprising:

\begin{itemize}
\item
A broad community discussion at a workshop to be held in Pasadena during 
13--16 June, 2000;
\item
Discussion among multiple community working groups identified at the 
Pasadena workshop and charged with developing:

  \begin{enumerate}
  \item
  The details of a major scientific program enabled by the technical 
  possibilities outlined above;
  \item
  An explanation of the scientific merit of the program as well as
  a discussion of the difficulty of accomplishing the same result without
  the NVO;
  \item
  An understanding of the flowdown from the needs of the science program 
  to archive, archive access, and software tool requirements; and
  \item
  Prioritization of requirements.
  \end{enumerate}

\item
A meeting among the chairs of the community working groups and the interim 
NVO steering committee to combine the input from the working groups into 
a coherent SRM, complete with a science-to-requirements flowdown for 
NVO.
\end{itemize} 

In preparation for this process, we have developed a few examples of 
science programs that give both a sense of the possibilities for discovery 
enabled by the NVO and of the initial basis for defining a framework and 
cadence for their implementation.  We caution that these examples are as
yet incomplete, and that they will require substantial community effort and
input to convincingly demonstrate the potential power of the NVO and its
required functions.  In the material presented here, we emphasize the
flowdown from science to capabilities because we believe that this step is
necessary to understand how a complete description of the NVO will be
obtained.

\vspace{0.5cm}

{\bf Example \#1:
A Panchromatic Census of Active Galactic Nuclei (AGN)} 
\vspace{0.3cm}

{\em Background:}  An understanding of the nature and characteristics of AGN is 
important both because their luminosities make them visible to large 
cosmological distances, and because they represent a fundamental stage in 
the evolution of galaxies.  Observationally, AGN are distinguishable from 
stars in that their spectral energy distributions are much broader than 
black-body functions.  However, redshift, variability, (possibly large) 
obscuration, and a range of intrinsic spectral shapes and characteristics 
result in a great range of ``colors'' over wavelength ranges from X-ray to 
radio.  

{\em Scientific Goals:}  This project aims to construct a complete sample of AGN in 
order to:
\begin{itemize}
\item
Test the idea that observable properties are determined by extrinsic factors 
such as orientation or obscuration (so-called unification models).
\item
Compare the environments of AGN as a function of type, \eg radio 
properties {\it vs}.\ membership in clusters of galaxies.
\item
Understand the evolution of the AGN luminosity function, and, in particular, 
separate number evolution from luminosity evolution.
\item
Construct the AGN luminosity function for different wavelengths in order to 
understand the evolution of AGN properties in a statistical sense.
\end{itemize}

{\em Outline of the Project:}
\begin{enumerate}
\item
Federate a number (N) of surveys covering the same (significant) area on the 
sky, and together, spanning a large wavelength range (X-ray through radio).
\item
Include metadata information so that survey selection effects and constraints 
can be accounted for.  Relevant metadata will include survey area, 
bandpasses, flux limits, \etc
\item
Identify distinguishable ``clouds'' of objects in N-dimensional color space in 
the resulting joint dataset.
\item
Apply a priori astronomical knowledge (\eg published catalogs, theoretical 
models) to understand population of these ``clouds''.
\item
Identify AGN candidates in the joint dataset.  Note that confirmation may 
require new observations the process might be carried out as a statistical 
one, resulting in a probability that any particular object is an AGN.
\item
Use catalog properties and/or further measurements from image databases to 
address science questions.
\end{enumerate}

{\em NVO Functionality required:}
\begin{itemize}
\item
Federation of relevant surveys including cross-identification of objects in 
multi-wavelength surveys and interchange/merging of metadata.
\item
Cluster analysis to identify ``clouds'' and ``sequences''.  This will include both 
supervised analysis (in which astronomical knowledge guides the definition 
and analysis) and unsupervised analysis (in which new patterns are 
recognized).
\item
Visualization of multi-dimensional datasets.
\item
Statistical analysis/classification of the populations of defined regions in the 
multi-dimensional parameter space.
\end{itemize}

{\bf Example \#2:
Formation and Evolution of Large-Scale Structure}
\vspace{0.3cm}

{\em Background:} Clusters of galaxies represent the largest unambiguous mass 
concentrations known.  Various models of the early evolution of the universe 
make different predictions for how clusters form and evolve.  These models can 
be tested by comparing their predictions with observed mass and luminosity 
functions of clusters.

{\em Scientific Goals:}  This project aims to construct an unbiased sample of clusters of 
galaxies over a cosmologically significant redshift range in order to test various 
structure formation and evolution models by comparison with evolution of the 
observed mass and luminosity functions with redshift.  In addition, the sample 
can be used to study the morphology-density relation for galaxies and its 
evolution over interesting timescales.

{\em Outline of the Project:}

\begin{enumerate}

\item Create statistical cluster samples using multi-wavelength pixel data in a 
number of different ways:

\begin{description}
\item
[X-ray surveys:] identify cluster signature from emission of hot gas in image data.
\item
[Optical/IR surveys:] convolution of image data with kernel designed to select 
clusters.
\item
[Millimeter surveys:] identify clusters from variation in CMB temperature caused 
by the Sunyaev-Zel'dovich effect.
\item
[Radio surveys:] identify clusters based on presence of radio source morphologies 
indicative of cluster environment.
\end{description}

\item Compare the results of different selection techniques.  Quantify the selection 
effects as functions of cluster mass, density, redshift, \etc

\item Use various distance indicators or redshift estimators to supplement 
measured properties of clusters.

\end{enumerate}

{\em Required NVO Functionality:}

\begin{itemize}
\item
Operate on large quantities of imaging data with user-defined algorithms and 
tools.
\item
Construct simulated surveys to understand selection effects; test user-defined 
tools on simulations.
\item
Generate predictions of observed sample properties based on various 
theories.  Compare with observed samples.
\end{itemize}

{\bf Example \#3:
The Digital Galaxy} 
\vspace{0.3cm}

{\em Background:} The Galaxy is composed of a number of structural elements: halo, 
thin disk, thick disk, bulge, spiral arms.  Each of these is characterized by 
populations of stars that have correlated distributions in age, mass, chemical 
composition, and orbit (position and kinematics), as well as distributions of non-
stellar material such as gas and dust.  These are the fossil tracers of the 
formation processes.  A complete understanding of these, in a global context, has 
never been possible because of the difficulty of studying samples of the size 
needed to disentangle all the variables simultaneously.  

{\em Scientific Goals:} This project aims to construct very large samples of 
galactic stars together with as much information about the physical 
properties of each object as can be derived.  These datasets, together 
with maps of non-stellar components of the Galaxy, will be used to:

\begin{itemize}
\item
Generate a parameterized model of the Galaxy, including positional and 
kinematic information.  
\item
Confront this model with models for the structure of the Galaxy, based on 
various formation processes.
\item
In particular, search for co-moving groups that are representative of merger 
events or tidal debris tails.
\end{itemize}

{\em Outline of the Project:}

\begin{enumerate}
\item Federate various optical and IR surveys to generate matched catalogs of 
stars.

\item Use positions, magnitudes, and colors to construct three-dimensional 
stellar distributions.  This will require using colors to derive 
luminosity classes and estimate extinction.

\item Quantify dust distribution and obscuration using FIR, H I, and CO images.

\item Iterate with 2) until consistent.

\item Identify bulk flows and sites of star formation using IR and radio images.

\item Use proper motion surveys (and radial velocity information, when available) 
to deduce motions of subsets of stars.

\item Use multi-epoch imaging to find variables.  Use these to provide a 
distance check.

\end{enumerate}

{\em Required NVO Functionality:}

\begin{itemize}
\item
Federation of multi-wavelength and multi-epoch catalog data.
\item
Operation on large quantities of image data with user-defined tools.
\item
Visualization tools for large multi-dimensional datasets.
\item
Statistical tools to analyze components and find coherent groups of objects.
\end{itemize} 

We emphasize again that these projects are meant to be illustrative of the
kind of  science, previously very difficult, that could be accomplished
using capabilities that  we foresee for the NVO.  As the NVO comes into
being, it is clear that more highly defined and diverse sets of projects
will be developed as a result of community input and discussion.
 
\section{Technical Issues}

\subsection{Overview}

In assessing the current state of North American astronomy, the
following resources are already in place to support the emerging NVO:

\begin{itemize}

\item {\em Data Centers and Supercomputer Centers.}
Some tens of Terabytes of data products (catalogs, images, and spectra)
already exist for various space missions, public telescopes, and surveys;
this will expand to a Petabyte or more of data by the end of the decade.
Archive and data analysis capabilities exist at the major NASA centers
(STScI, IPAC, HEASARC, and CXC) and at the CADC (Canada); many smaller or
more focused archives exist as well.  Supercomputer centers such as the SDSC
and NCSA are available for addressing large scale computational problems.
A high performance national networking infrastructure is already
in place.

\item {\em Astronomical Information Services.}
Information services such as the ADS, NED, and SIMBAD exist for name
resolution and cross-referencing of galactic and extragalactic objects,
and are providing increasingly sophisticated levels of interlinking
between bibliographic information, the refereed and preprint literature,
and the archival data centers.

\item {\em Data Analysis Software.}
Various software packages such as AIPS, AIPS++, IRAF, IDL, FTOOLS, SkyView,
\etc, exist for the general analysis of astronomical data. The development
of sophisticated software for large scale data mining is still in its
infancy, although new initiatives such as the NPACI-sponsored Digital Sky 
and the IPAC Infrared Science Archive are showing the potential of such 
facilities and have prototyped the technology required to correlate and 
mine such data archives.

\end{itemize}

Although these resources are significant, anyone who has tried to perform
multiwavelength data analysis or large scale statistical studies combining
several different catalogs, with the data involved being available from
widely distributed and dissimilar archives, will appreciate how far we have
to go to implement the vision of the NVO.
Ground-based O/IR and radio data need to be pipeline-processed and archived
routinely as space-based data are now.  Standards and protocols need to
be developed to allow widely distributed archives to interoperate and
exchange data.  Astronomical data analysis software needs to evolve to
be able to access data in distributed multiwavelength archives as easily
as local datasets are accessed now.  New algorithms, applications, and
toolkits need to be developed to mine multi-Terabyte data archives.
Supercomputer-class computational systems need to be developed to enable
large scale statistical studies of massive, multiwavelength distributed
data archives.  The data, software, and computational resources need to 
be interconnected at the highest available network bandwidths.

\subsubsection{Data Archives}

Any consideration of the science to be performed by the NVO, or the
technical issues involved in implementing the NVO, must start with
the {\em data}.  Although the data from most NASA missions have been
routinely archived for over a decade, relatively little data from ground
based telescopes is currently available online, other than for a few
major surveys.  With modern wide-field and multispectra instruments on
ground-based telescopes producing ever larger quantities of data, and
with ground-based survey projects becoming almost as common as classical
observing, {\em there is an acute need to archive and publish high quality
datasets from ground-based instruments and surveys}.  The science promised
by the NVO will not be possible unless the NVO succeeds in creating true,
panchromatic images and catalogs, seamlessly integrating
data from both ground- and space-based archives, and thereby enabling
exploration of astrophysical phenomena over most of the electromagnetic
spectrum.

Experience over the past decade has shown that astronomical archives
are complex and diverse, never stop growing, and are best maintained by
those close to the data who know it well.  In practice this has meant that
most data are put online either by individual large survey projects, \eg
the 2-Micron All-Sky Survey (2MASS) or the Sloan Digital Sky Survey (SDSS), 
or by discipline specific archive centers
which serve a given community.  To address the need to move to large
scale archiving of ground-based astronomical data, archiving facilities
will need to be established at the national centers (NOAO, NRAO, NSO,
NAIC), and partnerships will need to be formed with the major private
and university-operated facilities.  The major national data centers
for ground- and space-based data will comprise the principle nodes of the
distributed NVO data system in the U.S.

\subsubsection{Technical Challenges}

Given archival quality data from all branches of astronomy, physically
distributed at 10--20 major archive centers and any number of ancillary
datasets together with a distributed community of thousands of scientific 
users, one can define what new software and services will be required 
to implement the NVO.  Analysis of
the data will be complex, due to the heterogeneous nature of datasets
from the different branches of astronomy and due to the use of increasingly
complex data structures (within the general framework of the FITS data
format standard) to accommodate the increasing levels of sophistication
of modern astronomical instrumentation.

The sheer scale of the problem is daunting, 
with catalog sizes approaching the Terabyte
range and the total data volume in the Petabyte range.  However, an
{\em even more serious challenge comes from the complexity of these
datasets,} with tens or hundreds of attributes being measured for
each of ten million or more objects.  This is {\em a crucial new aspect
to the data mining issue,} and multivariate
correlation of such large catalogs is a massive computational problem.
If pixel level analysis of candidate objects is required the computational
problem can be even more massive.  It is important to recognize that 
current brute-force analysis techniques do not scale to problems of this
size!  Multidisciplinary research in areas such as metadata representation
and handling, large scale statistical analysis and correlations, and
distributed parallel computational techniques will be required to address
the unprecedented data access and computational problems faced by the NVO.

Fortunately, astronomy is not alone in facing this problem.
{\em The technological challenges for the NVO are similar to those facing other
branches of science,} such as high energy physics, computational genomics,
global climate studies, and oceanography.  Research and development of
information systems technology is already underway in areas such as
statistical analysis and data mining of large archives, distributed
computational grids, data intensive grid computing (data grids),
and management of structured digital information (digital libraries).
Much of this research is relevant to the problems faced by the NVO.
Information technology and data management throughout the sciences will
both advance, and be advanced by, the NVO.

The large dataset size and geographic distribution of users and resources
also presents major challenges in connectivity.  Next generation networking
providing cross-continental bandwidths of 100 MB/sec is now available
and currently underutilized, but this situation will change rapidly.
It will be essential for the major NVO data centers to be interconnected
with very high speed networks, and to utilize intelligent server-side
software agents in order to make the most efficient use of the network
when interacting with end-users.

\subsection{Architecture}

The technical challenge of implementing the NVO is a study in contrasts.
While data will be widely distributed, the large studies
at the cutting edge of the science to be enabled by the NVO will need
massive computational resources and fast local access to data.  While
sophisticated metadata standards and access protocols will be required
to link together distributed archives and network services, the effort
required to interface a small archive to the NVO must be minimized to
encourage publication of new data collections by the community.  While data
collections and compute services will be widely distributed, users will
need a straightforward interface to the system which makes the location
and storage representation of data and services as transparent as possible.

To meet this wide range of requirements, the NVO needs a distributed system
architecture that provides uniform and efficient access to data and services
irrespective of location or implementation.  {\em Data archives} are assumed
to already exist and will vary considerably in implementation and access
policy.  {\em Metadata standards} will be devised to provide a well defined
means to describe archives, data collections, and services.  A {\em data
access layer} will provide a single uniform interface to all data and
services, and will be used both to link archives and services within the
framework of NVO, and to allow user applications to access NVO resources.
{\em Query and compute services} will provide the tools for information
discovery and large scale correlation and analysis of disparate datasets.
{\em Data mining applications}, running on a user workstation at their home
institution, as applets within a Web browser, or at a major NVO data center,
will provide the main user interface to enable science with the NVO.

\subsection{Components}

\subsubsection{Data Archives}

Data archives store {\em datasets} (\eg catalogs, images, and spectra)
organized into logically related {\em data collections}, as well as
{\em metadata} describing the archive and its data holdings.  Access is
provided in various ways such as via a structured Web interface, via a
standard file-oriented interface such as FTP, or via other access protocols
which may vary from archive to archive.

NVO will place no requirements on data archives other than that they be
made accessible to NVO via the data access layer (DAL), which serves as
the portal by which NVO gains access to the archive.  In the simplest
cases interfacing an archive to NVO will be little more than a matter of
installing the data access layer software and modifying a few configuration 
files 
to reflect the data holdings and access permissions of the local archive,
much as one would install a Web server.  More sophisticated installations
may provide expanded support for metadata access and server-side functions,
as outlined in the discussion of the data access layer below.

\subsubsection{Metadata Standards}

Metadata (literally, ``data about data'') is structured information
describing some element of the NVO.  Metadata will be required to
describe archives, the services provided by those archives, the data
collections available from an archive, the structure and semantics
(meaning) of individual data collections, and the structure and semantics
of individual datasets within a collection.  Typical astronomical datasets
are data objects such as catalogs, images, or spectra.  As an example,
the semantic metadata for a typical astronomical image is the logical
content of the FITS header of the image.

Metadata describing astronomical data is essential to enable {\it data
discovery} and {\it data interoperability}.  Metadata describing archives and
services is necessary to allow the components of NVO to interoperate in an
automated fashion.  {\it Metadata standards} are desirable to make these
problems more tractable.  In practice there are limits to what can be done to
standardize dataset specific metadata, but mediation techniques such as those
being developed by the digital library community provide ways to combine
metadata dialects developed by different communities for similar types of
data.  Current projects such as Astrobrowse and ISAIA (Interoperable 
Systems for Archival Information Access) represent initial efforts within
the astronomical community to establish metadata standards.

\subsubsection{Data Access Layer}

The data access layer (DAL) will provide a uniform interface to all data,
metadata, and compute services within NVO.  At the lowest level the data
access layer is a standard {\em protocol} defining how the software
components of the NVO talk to each other.  Reference grade software
implementing the protocol will also be provided, which can either be used
directly or taken as the basis for further development by the community.
This software will include server-side software used to interface archives
and compute services to the NVO, and client-side applications programming
interfaces (APIs), which can be used to write NVO-aware distributed
data mining applications.  Since the DAL is fundamentally a protocol,
multiple APIs will be possible, \eg to support legacy software or multiple
language environments.

The key aspect of the data access layer is that it provides a uniform 
interface to {\em all} data and services within NVO. User applications 
use the data access layer to access NVO data and services, and archives
and compute services {\em within} the NVO use the data access layer 
internally to access data or services in other archives, potentially 
generating a cascade of such references.  NVO is thus an inherently 
hierarchical, distributed system, which nonetheless has a simple structure 
since all components share the same interface.  In addition to such {\em 
location transparency}, the data access layer will provide {\em storage 
transparency}, hiding the details of how data are stored within an archive.  
Finally, the data access layer protocol will define standard data models 
(at the protocol level) for astronomical data objects such as images 
and spectra.  Archive maintainers will provide server-side modules to 
perform {\em data model translation} when data objects are accessed,
allowing applications to process remote data regardless of its source or
how it is stored within a particular archive.

Often a client program using the data access layer will not need an entire 
dataset, but only a portion.  {\em Server-side functions} will permit 
subsetting, filtering, and data model translation of individual datasets.  
In some cases {\em user defined functions} may be downloaded and applied 
to the data to compute the result returned to the remote client.  This is
critical to reduce network loading and distribute computation.

Since the data access layer can be used to read both metadata and actual 
datasets from a remote archive, {\em dataset replication} becomes possible, 
allowing a {\em local data cache} to be maintained.  This is critical to 
optimizing data access throughout the NVO, and will be necessary to even 
attempt many large scale statistical studies and correlations.  Dataset 
replication also makes it possible to replicate entire data collections, and to migrate
data archives forward in time.  Metadata replication and ingest makes it
possible for a central site to automatically index entire remote archives.

\subsubsection{Query and Compute Services}

While the data access layer and metadata standards will allow the NVO
to link archives and access data, {\em query and compute services} will
be required to support information discovery and provide the statistical
correlation and image analysis capabilities required for data mining.

While most archives will provide basic query services for the data
collections they support, large scale data mining does not become possible
until multiple catalogs are combined (correlated) to search for objects
matching some statistical signature.  The larger NVO data centers will
provide the data and computational resources required to support such
{\em large scale correlations}.  While a query or correlation may result
in subqueries to remote archives, extensive use of dataset replication
and caching will be employed to optimize queries for commonly accessed
catalogs or archives.  Sophisticated metadata mediation techniques will be
required to combine the results from different catalogs.

In some cases pixel-level analysis of the original processed data, using an
algorithmic function downloaded by the user, may be required to compute new
object parameters to refine a parametric search (in effect this operation
is dynamically adding columns to an existing catalog, an extremely powerful
technique).  Since with NVO candidate object lists may contain several
hundred million objects, this is a massively parallel problem such as
might require a Terascale supercomputer to address.  Even in the case
of large scale statistical studies, distributed computing techniques and
fast networks may allow the user to work from their home institution, but
some form of peer reviewed time allocation may be required to allocate the
necessary computational and storage resources.  For some larger studies
users may need to visit a NVO data center in order to have efficient
access to personnel as well as data, software, and computational resources.

\subsubsection{Data Mining Applications}

The field of data mining, including visualization and statistical analysis
of large multivariate datasets, is still in its infancy.  This will
be an area of active research for many years to come.  Most current
astronomical data analysis software will need to be upgraded to become
``NVO-aware'', able to be used equally well on both local and remote data.
New applications will be developed as part of ongoing research into data
mining techniques.  While NVO should provide the interfaces and toolkits
required to support this development, as well as some initial data mining
applications from the major NVO centers, the open-ended nature of the
problem suggests the need for a multidisciplinary data mining research
grants program once NVO becomes operational.

\subsubsection{Information Systems Research}

In all areas---storage technology, information management, data
handling, distributed and parallel computing, high speed networking,
data visualization, data mining---NVO will push the limits of current
technology.  Partnerships with academia and industry will be necessary
to research and develop the informations systems technology necessary
to implement NVO.  Collaborations with other branches of science and
with the national supercomputer centers will be required to develop
standards for metadata handling, data handling and distributed computing.
Data mining is an inherently multidisciplinary problem which will require
the partnership of astronomers, computer scientists, mathematicians,
and software professionals to address.

A next generation, high speed national research Internet is already
in place, but is underutilized at present due to the lack of credible
academic applications designed to make use of high performance networking.
NVO would be a prime example of a creative new way to use wide area high
performance networking for academic research.

\subsubsection{Education and Public Outreach}

Given the wealth of real science data the NVO will make freely available
via the Internet, and the keen interest of the public in astronomy, 
{\em the NVO will be uniquely suited 
for education and advancing science literacy.}
The intrinsic Internet-based nature of the NVO lends itself
to a variety of high-quality science popularization and education methods
with {\em an unprecedented social and geographical outreach.}

We anticipate that education and outreach professionals (educators,
staff members of planetaria, science museums, popular
science writers and journalists, \etc) would become actively involved in
utilizing this remarkable set of resources, creating of popular science
websites, course materials (from elementary to graduate school), and
sophisticated demonstrations.  Schools with modest science education
resources would be able to find hands-on demonstrations on line.
Applet software running in commodity web browsers will permit NVO data to
be accessed and visualized by the public, allowing virtual observations
to be taken and the resultant data analyzed and interpreted.
We expect that {\em a range of outreach
partnerships will be developed with the NVO as a centerpiece and as a
catalyst.}

The NVO is especially interesting as a science and technology education
focus because {\em it bridges a physical science (astronomy) and applied
computer science.}  It thus employs a range of technologies and skills
relevant  to many aspects of the economy and society as a whole in the 21st
century. Real-life examples of the use of such methodologies can be a
powerful way to teach material that may otherwise be a very dry or
difficult.  For example, we note the great popular success of the SETI@home
project; one can envision more sophisticated examples where data mining
techniques are  used by large numbers of people on comparably exciting
problems spawned within the NVO.

\section{Implementation Plan}

Previous sections have described the technological changes that will
enable a ``new astronomy'' and the characteristics of an NVO that
can capitalize and build upon those changes to enable new and more cost
effective science than would otherwise be possible.

The fundamental basis for the NVO management activities will be to
recognize the science driven nature of the NVO
and to maximize the community participation in the NVO
effort. There will be three levels of activity and funding. These are
structured to ensure that there is a usable and well documented
infrastructure, that the software projects are science driven, and
that bulk of the funding is dispersed to well focused science based
proposals that are peer reviewed.

\begin{enumerate}

\item The highest priority is to build the archive infrastructure and
well documented protocols to access the data. These will be standards
for data access that the community can rely on to build higher level
tools. These would evolve as the technology advances, but should
always be backward compatible. This infrastructure is funded via a
base budget and is developed and maintained by the major NVO distributed
sites.

\item There will be regular ``AOs'' for opportunities to build ``software
tools'' that utilize the infrastructure. They would be delivered to
the NVO for wider use by the community and would follow standards
defined by the NVO. It is important that these tool
building opportunities cover a wide range of possibilities and engage
a large part of the community.
A strong science enabling case for each software tool must be
made, but they will be general user facilities that the entire
community can use to do research.

\item There are regular ``AOs'' to use the NVO. These would be more
specific research projects with a well defined goal, that might
include software development. (This would be similar to the current
NASA ADP program). These would be much less structured in the sense
of being grants and with the deliverable being a paper to a journal.

\end{enumerate}

In the early phases of the NVO, the emphasis may be on the first two
areas, but as the NVO infrastructure develops the balance
of the funding between these three areas will evolve.

Implementation of the NVO would occur in several stages, from preliminary
preparation to fully operational stages.  {\em A major objective of
the implementation plan is to begin providing some levels of functionality
as quickly as possible through use of existing tools and services}.

\subsection{Phase 0: Prior to NVO Start}

\textit{Goal: Create the conceptual design of the NVO; begin activities at some
centers to provide necessary capability for implementation of the NVO.}
\begin{itemize}
\item
Develop relevant position papers, supporting documents, and a ``Science
Reference Mission'' which identifies the key science goals for the NVO;
\item
Initiate work within participating organizations to ensure accessibility
of data and the establishment of archives;
\item
Develop essential enabling technologies, such as information exchange
protocols and metadata standards;
\item
Establish catalog search and/or image data retrieval capability for
selected data subsets at all major sites;
\item Initiate community involvement through meetings and workshops; and
\item Open lines of communication with the international community
concerning general NVO initiative.
\end{itemize}

\subsection{Phase 1: Months 1--18}

\textit{Goal: Establish integrated data discovery, data delivery,
and data comparison services.}
\begin{itemize}
\item
Expand and formalize the data discovery and data delivery systems,
including establishment of metadata standards, transport protocols,
and presentation services;
\item
Continue to work with all sites to improve access to online services;
\item
Plan for eventual network connectivity and computational requirements;
\item
Deploy small scale cross-correlation capabilities and visualization
tools;
\item
Prototype large scale cross-correlation facilities;
\item
Continue community involvement through meetings, workshops, and the
establishment of a Users' Committee and a Visiting Committee;
\item
%Identify Program Manager and Project Scientist,
Establish core technology and management groups and establish reporting
and accountability procedures;
\item
Identify subsets of NVO functionality that can be most effectively
developed by existing data centers or other entities;
\item
Establish an outreach program;
\item
Move forward in the design and establishment of an international information
infrastructure for astronomy; and
\item
Foster communication and collaborations with efforts to advance
information technology in other fields.
\end{itemize}

\subsection{Phase 2:  Months 18--36}

\textit{Goal: Establish initial large scale cross-correlation
capabilities; begin full scale operations.}
\begin{itemize}
\item
Begin putting network and associated computing facilities in place;
\item
Develop and deploy visualization tools for complex datasets;
\item
Develop and make initial deployment of the data access layer (DAL);
\item
Ensure that the data discovery and comparison tools are now mature
and in routine operation;
\item Establish partnerships with international organizations to assure
interoperability of US and non-US facilities and services; 
\item
Management structure and advisory committees now in routine operation.
\end{itemize}

\subsection{Phase 3: Months 36--60}

\textit{Goal: Establish fully operational baseline NVO; enable full
scale cross-correlations supported by suitably configured computational
and network systems.}

\begin{itemize}
\item
Breadth of data services extended to additional facilities, including
international collaborators;
\item
Data access layer deployed and in routine operation;
\item
Support user-defined portable processing agents;
\item
Establish support of higher level data products, such as pre-prepared
cross identifications.
\end{itemize}

This implementation timetable is only approximate and will naturally
evolve as the problems become more well defined and as the level of
support for the NVO becomes clearer.  However, this timetable is
``optimal'' in that it reflects estimates of an ideal implementation
path for the NVO functionality.  Services are estimated to be implemented
as rapidly as is feasible from a technical point of view; restricted
funding levels below the optimal level would clearly slow this process.

\end{document}